%Paper: hep-th/9512137
%From: RCM@hep.Physics.McGill.CA
%Date: Mon, 18 Dec 1995 16:48:35 -0500 (EST)

\input harvmac

%%%%%%%%%%%%%%%%%%
\Title{\vbox{\baselineskip12pt
\hbox{CERN-TH/95-345}
\hbox{McGill/95-25}
\hbox{hep-th/9512137}
}}
{\vbox{\centerline{Low-Energy Scattering of Black Holes and}
\smallskip     \centerline{$p$-branes in String Theory}}}

\centerline{Ramzi R.~Khuri$^{a,b}$ \foot{Talk given by R. Khuri at the
Sixth Canadian Conference on General Relativity and
Relativistic Astrophysics, University of
New Brunswick, Fredericton, N.B., Canada, May
1995, and to appear in Proceedings thereof.}
 and Robert C.~Myers$^b$}
\bigskip\centerline{$^a${\it CERN,
 CH-1211, Geneva 23, Switzerland}}
\bigskip\centerline{$^b${\it Physics Department, McGill
University, Montreal, PQ, H3A 2T8 Canada}}
\vskip .3in
We discuss the low-energy dynamics of generalized
extremal higher membrane black hole solutions of string theory
and higher membrane theories following Manton's prescription
for multi-soliton solutions.
A flat metric is found for those solutions which possess
$\kappa$-symmetry on the worldvolume.

\vskip .3in
\Date{\vbox{\baselineskip12pt
\hbox{CERN-TH/95-345}
\hbox{McGill/95-25}
\hbox{December 1995}}}

\def\sqr#1#2{{\vbox{\hrule height.#2pt\hbox{\vrule width
.#2pt height#1pt \kern#1pt\vrule width.#2pt}\hrule height.#2pt}}}

\lref\stst{M.~J.~Duff and R.~R.~Khuri, Nucl. Phys. {\bf B411}
(1994) 473; M.~J.~Duff, Nucl. Phys. {\bf B442} (1995) 47.}

\lref\prep{M.~J.~Duff, R.~R.~Khuri and J.~X.~Lu,
Phys. Rep. {\bf 259} (1995) 213.}

\lref\dynam{R.~R.~Khuri and R.~C.~Myers,
``Dynamics of extreme black holes and massive string states'',
preprint hep-th/9508045 (to appear in Phys. Rev. D)}

\lref\rusty{R.~R.~Khuri and R.~C.~Myers,
``Rusty scatter branes'', preprint hep-th/9512061.}

\lref\dufr{M.~J.~Duff and J.~Rahmfeld, Phys. Lett.
{\bf B345} (1995) 441.}

\lref\dufkmr{M.~J.~Duff, R.~R.~Khuri, R.~Minasian and
J. Rahmfeld, Nucl. Phys. {\bf B418} (1994) 195.}

\lref\monscat{R.~R.~Khuri, Phys. Lett. {\bf B294} (1992) 331.}

\lref\stscat{R.~R.~Khuri, Phys. Lett. {\bf B307} (1993) 302.}

\lref\fbscat{A.~G.~Felce and T.~M.~Samols, Phys. Lett.
{\bf B308} (1993) 302.}

\lref\macstr{R.~R.~Khuri, Nucl. Phys. {\bf B403} (1993) 335.}

\lref\hmono{R.~R.~Khuri, Phys. Lett. {\bf B259} (1991) 261;
Nucl. Phys. {\bf B387} (1992) 315.}

\lref\manton{N.~S.~Manton, Phys. Lett. {\bf B110} (1982) 54;
Phys. Lett. {\bf B154} (1985) 397;
Phys. Lett. {\bf B157} (1985) 475.}

\lref\atiyah{M.~F.~Atiyah and N.~J.~Hitchin, Phys. Lett.
{\bf A107} (1985) 21; {\it The Geometry and Dynamics of
Magnetic Monopoles}, (Princeton University Press, 1988).}

\lref\fere{R.~C.~Ferrell and D.~M.~Eardley,
Phys. Rev. Lett. {\bf 59} (1987) 1617;
in {\it Frontiers in numerical relativity,}
Eds. C.R.~Evans, L.S.~Finn and D.W.~Hobill
(Cambridge University Press, 1988).}

\lref\ghs{D.~Garfinkle, G.~T.~Horowitz and A.~Strominger,
Phys. Rev. {\bf D43} (1991) 3140.}

\lref\stata{A.~Papapetrou, Proc. R. Irish Acad. {\bf A51}
(1947) 191; S.D. Majumdar, Phys. Rev. {\bf D35} (1947) 930.}

\lref\salam{S.~W.~Hawking, Monthly Notices Roy. Astron. Soc.
 {\bf 152} (1971) 75; A.~Salam in {\it Quantum Gravity: an
Oxford Symposium} (Eds. Isham, Penrose and Sciama, O.U.P. 1975);
G.~'t Hooft, Nucl. Phys. {\bf B335} (1990) 138.}

\lref\shirscat{K.~Shiraishi, Nucl. Phys. {\bf B402} (1993) 399;
``Many-body systems in Einstein-Maxwell-Dilaton theory,''
preprint gr-qc/9507029.}

\lref\shir{K.~Shiraishi, J. Math. Phys. {\bf 34} (1993) 1480.}

\lref\gibbons{G.~W.~Gibbons, Nucl. Phys. {\bf B207} (1982) 337.}

\lref\calscat{C.G.~Callan, J.M.~Maldacena and A.W.~Peet,
``Extremal Black Holes as Fundamental Strings,'' preprint hep-th/9510134;
A.~Dabholkar, J.P.~Gauntlett, J.A.~Harvey and D.~Waldram,
``Strings as Solitons and Black Holes as Strings,''
preprint hep-th/9511053;
C.F.E.~Holzhey and F.~Wilczek,
Nucl. Phys. {\bf B380} (1992) 447;
F.~Larsen and F.~Wilczek, ``Internal Structure of Black Holes,''
preprint hep-th/9511064;
G.~Mandal and S.R.~Wadia, ``Black Hole Geometry
around an Elementary BPS String State,''
preprint hep-th/9511218.}

%%%%%%%%%%%%%%%%%%%%%%%%%%%%%%%%%%%%%%%%%%%%%%%%%%%%%%%%%%%%%%%%

\newsec{Introduction}

The construction of soliton and black hole solutions of string
theory and their connections with various dualities in
string theory have been the subject of much recent activity
(see \prep\ and references therein).
The soliton solutions typically arise as extremal limits of
two-parameter charged black hole solutions. These extremal
black holes saturate a Bogomol'nyi bound between their
ADM mass and charge, thus ensuring their stability, in analogy
with the extremal Reissner-Nordstrom black holes of
Einstein-Maxwell theory. In both the string and
Einstein-Maxwell cases, the saturation of this
bound is associated with the existence of a ``zero-force''
condition, which allows for the existence of multiple
extreme black holes in a static configuration. In the string
context, the saturation of the Bogomol'nyi bound is also
associated with the existence of spacetime supersymmetry.
A further generalization inherent in the string scenario is
that these black hole solutions may have higher membrane
structure, arising in a higher-dimensional spacetime.

Most of the solutions discussed in the string context
are associated with an underlying ``worldvolume'' supersymmetry
of a higher-membrane theory (generalizing worldsheet
supersymmetry in string theory),
called $\kappa$-symmetry, which ensures that the correct number
of degrees of freedom arise in the theory. It turns out that
these $\kappa$-symmetric solutions are also associated
with the preservation the maximal amount of spacetime
supersymmetry in various embeddings.

In this work, we start with the generalized extremal black
holes of Shiraishi \shir, and note
that these include both $\kappa$-symmetric and
non-$\kappa$-symmetric solutions. We then generalize these
solutions further to arbitrary
higher-membrane black holes embedded in arbitrary higher
dimensions and discuss their connections with the various
dualities in string and higher membrane theories.

In section 3 we follow Manton's method \manton\
for the computation of the low-velocity scattering of both
the $\kappa$-symmetric and non-$\kappa$-symmetric
generalized extreme higher-membrane black holes and find that
this scattering is trivial ({\it i.e.,} zero dynamical force)
precisely for the $\kappa$-symmetric solutions.

Finally, we discuss our results in section 4 as well
as the intriguing possibility that some of the extremal
black holes we discuss might correspond to elementary
states in string theory.

\newsec{Generalized extreme black hole solutions}

Consider the action
\eqn\action{
I_4 (1) = \int d^4 x \sqrt{-g}
\Bigg(R - {1\over 2} (\partial\phi)^2 -
{1\over 4} e^{-a\phi} F^2_2\Bigg),}
where $g_{MN}$ is the metric, $\phi$ the scalar dilaton
and $F_2= dA_1$ the Maxwell field strength.
$a$ is an arbitrary constant.
The generalized multi-extreme black hole solution is given
by \shir
\eqn\solution{\eqalign{
ds^2&=-F(x)^{-2/(1+a^2)} dt^2 + F(x)^{2/(1+a^2)} d\vec x \cdot
d\vec x ,\cr
\phi&=-{2a\over 1+a^2} \ln F,\cr
[A_1]_0&= \pm \sqrt{{4\over 1+a^2}} F, \cr
F(x)&= 1+ \sum_{i=1}^N {k_i\over |\vec x - \vec x_i|},\cr}}
where $\vec x_i$ are the locations of the $N$ black holes.
The mass $m_i=k_i/(1+a^2)$ and electric charge
$e_i=2k_i/\sqrt{1+a^2} $ of each black hole saturate the
Bogomol'nyi
bound $m_i^2\geq e_i^2/4(1+a^2)$, thus ensuring the stability
of the multi-black hole configuration.

The single black hole
solution represents the extremal limit of a two parameter
family of charged black hole solutions in which this bound
is not saturated. One can equally well find magnetically
charged solutions, where the magnetic charge is related to the
electric charge via the Dirac quantization condition
$eg=2\pi n$, where $n$ is an integer. For the specific value
of $a=\sqrt{3}$, the action couples to a sigma-model of
a theory with a fermionic worldline symmetry called
$\kappa$-symmetry,
which ensures that the nonphysical degrees of freedom in
the worldline theory are
projected out (see \prep\ and
references therein).

It is possible to generalize the above solutions in two
ways. In the first step, we can consider point-like black
holes in arbitrary higher-dimensional spacetime $D$. The
four-dimensional action \action\ is then replaced by a
$D$-dimensional action via $\int d^4x\to \int d^Dx$ and
$I_4(1)\to I_D(1)$. Note that the argument $1$ in the action
represents the dimension of the worldline swept out by the
point-like black holes. In this case, the $\kappa$-symmetric
solutions correspond to $a^2=2(D-1)/(D-2)$.

Furthermore\foot{The procedure is only outlined below. For
details see \rusty.},
we can generalize the point-like black holes themselves to
higher-membrane objects, with, say, $d-1$ spatial directions
(called a $(d-1)$-brane) by coupling to an antisymmetric
tensor $A_d$ with $d$ indices. Here the replacement takes
the form $(1/4)F_2^2\to (1/2(d!))F_{d+1}^2$, where
$F_{d+1}=dA_d$ is the field strength associated with $A_d$
in further generalizing the action via
$I_D(1)\to I_D(d)$.

A useful dimension to define is the ``dual'' dimension
$\tilde{d} \equiv D - d - 2$. Then for a $(d-1)$-brane in $D$
dimensions with ``electric'' charge $e_d$ arising from the
antisymmetric tensor, the natural solitonic object is
a $(\tilde d -1)$-brane carrying  topological ``magnetic''
charge
$g_{\tilde{d}}$ \prep. These again
are related by a (higher-dimensional) Dirac quantization condition.
In this
most general case, the $\kappa$-symmetric solutions correspond
to $a^2=4-2d\tilde d/(D-2)$.

The existence of these generalized dual solitons is in
fact the basis for the various dualities in string and
higher-membrane theories. An especially interesting example
of a duality which is of current interest is string/string
duality in $D=6$ ($d=\tilde d=2$). An interesting consequence
of this duality on reduction to four dimensions
is the interchange of two previously studied
dualities: the target-space $T$ duality, which generalizes
the compactification scale size duality in string theory,
and the strong/weak coupling $S$ duality \stst. An important
implication of the latter lies in the application of
perturbative techniques in the strong coupling region of string
theory, which is of special interest in the attempt to
understand string theory as a theory of quantum gravity.
Furthermore, the existence of these dualities is likely to
point to a reformulation of string theory
in which these dualities are manifest.

\newsec{Metric on moduli space}

In the absence of exact solutions for time-dependent
multi-soliton or multi-extreme black hole solutions,
Manton's method  for the computation of
the metric on moduli space for two or more solitons
represents a good low-velocity approximation to the
exact dynamics.

Manton's prescription for the study of soliton scattering
may be summarized as follows: One begins with a static
multi-soliton solution, and gives the moduli characterizing
this configuration a time-dependence. One then finds $O(v)$
corrections to the fields by solving the constraint
equations of the system with time-dependent moduli.
The resultant time-dependent field configuration only
satisfies the full time-dependent field equations to lowest
order in the velocities, but provides
an initial data point for the fields and their time derivatives.
Another way of saying this is that the initial motion is
tangent to the set of exact static solutions. An effective
action
describing the motion of the solitons is determined by
replacing the solution to the constraints into the field theory
action. The kinetic action so obtained
defines a metric on the moduli space of static solutions, and
the geodesic motion on this metric determines the dynamics
of the solitons \manton.

A calculation of the metric on moduli space for the scattering
of extreme Reissner-Nordstrom black holes and a description of
its geodesics was worked
out by Ferrell and Eardley \fere.
Shiraishi \shirscat\ then extended this work
to black hole solutions of \action\ in four dimensions
and of $I_D(1)$ in $D$ dimensions, all for arbitrary $a$.

Following Ferrell and Eardley's technique, it is possible to
compute this metric for the generalized, higher-membrane
 extreme black holes
of the previous section \rusty.
It turns out that precisely for
the $\kappa$-symmetric solutions, this metric is flat, and
the kinetic Lagrangian describes noninteracting extreme black
holes.

For the more general case, the direct computation of the metric
is also possible, but rather tedious. A tremendous
simplification occurs when one realizes that the low-velocity
dynamics of an arbitrary $(d-1)$-brane extremal black hole
embedded in $D$ dimensions are simply related to that of the
dimensionally reduced point-like black hole in $D-d+1$
dimensions \rusty, essentially due to
the fact that in both cases the dual dimension is the same
and given by,
$\tilde d=D-d-2$. This can be seen schematically via
\eqn\reduction{
(D,d) \to (D-1, d-1) \to \ldots \to (D-d+1, 1),}
where the first number in braces
indicates the spacetime dimension and
the second the the dimension of the worldvolume swept out by
the higher-membrane black hole,
and where the arrows indicate dimensional reduction of both
dimensions. Notice that the last pair of dimensions represents
a point-like black hole.

As the dynamics of the point-like
solutions have already been worked out by Shiraishi
\shirscat, we obtain our answer quite easily,
taking care to account for the proper volume terms for the
higher-membrane black holes.
Note that, in particular, the $\kappa$-symmetric solutions
can be expanded or reduced only to other $\kappa$-symmetric
solutions, since only these solutions scatter trivially.

As an example of nontrivial scattering for non-$\kappa$-symmetric
extreme black holes, consider the case of two point-like black
holes in $D=4$ with $a=1$ in \action. Then the low-velocity
Lagrangian is given following Manton's method by \shirscat
\eqn\kinlag{
L=-M+ {1\over 2} M V^2 + {1\over 2} \mu v^2\left(1+{2M\over r}
\right),}
where $M=m_1+m_2$ is the total mass,
$\vec V=\vec v_1+\vec v_2$ is the
center-of-mass velocity, $\mu=m_1m_2/M$ is the reduced mass,
$\vec v=\vec v_2 - \vec v_1$ is the relative velocity and
$\vec r = \vec x_2 -\vec x_1$ is the relative separation of the
two black holes. This interaction in fact yields
Rutherford scattering.

\newsec{Discussion}

The flat metric and consequent trivial dynamics for the
$\kappa$-symmetric solutions is a somewhat surprising result,
and is probably connected with the existence of flat directions
in the superpotentials associated with the underlying
$\kappa$-symmetric theories. Another possibility is that
these solutions also possess the maximal amount of spacetime
supersymmetry in certain embeddings, and this may also
constrain the dynamics considerably. For example,
if we embed the four-dimensional black holes in $N=8$,
$D=4$ supergravity, only the $\kappa$-symmetric
$a=\sqrt{3}$ black hole preserves four of the spacetime
supersymmetries.

Another interesting idea connected with the dynamics of these
solutions is the possibility that the point-like
extremal black holes might correspond to massive elementary
states in the string spectrum. Particles with mass
greater than the Planck mass possess a Schwarzschild radius
greater than their Compton wavelength, and so presumably
may display an event horizon even in a quantum-mechanical
framework. In the absence of a theory of
quantum gravity, however, this notion is hard to test. Since
string theory is a possible theory of quantum gravity and
predicts the existence of massive states at the Planck scale,
it provides a framework in which to test this conjecture.
Duff and Rahmfeld \dufr\
argued that certain massive string states in fact correspond
to extremal black hole solutions of \action.
More recently, dynamical evidence for this conjecture
was found by the present authors \dynam
(see also \calscat),
where the dynamics of the black holes according to the
Manton method and as discussed in this paper were compared
with the string-theoretic scattering amplitudes of the
corresponding string states. Remarkably, these two
seemingly disparate methods yield the same long-range
scattering. These results then promise to shed light on the
nature of string theory as a theory of quantum gravity.

\vskip1truecm

\noindent
{\bf Acknowledgements:}

\noindent
This research was supported by NSERC of Canada and Fonds FCAR du Qu\'ebec.
R.K. was supported by a World Laboratory Fellowship.

\vfil\eject
\listrefs
\bye